\documentstyle[prl,aps,floats]{revtex}

\begin{document}
\preprint{hep-ph/9912473}
\draft

%
%
\input epsf
\renewcommand\({\left(}
\renewcommand\){\right)}
\renewcommand\[{\left[}
\renewcommand\]{\right]}

\newcommand\del{{\mbox {\boldmath $\nabla$}}}

\newcommand\eq[1]{Eq.~(\ref{#1})}
\newcommand\eqs[2]{Eqs.~(\ref{#1}) and (\ref{#2})}
\newcommand\eqss[3]{Eqs.~(\ref{#1}), (\ref{#2}) and (\ref{#3})}
\newcommand\eqsss[4]{Eqs.~(\ref{#1}), (\ref{#2}), (\ref{#3})
and (\ref{#4})}
\newcommand\eqssss[5]{Eqs.~(\ref{#1}), (\ref{#2}), (\ref{#3}),
(\ref{#4}) and (\ref{#5})}
\newcommand\eqst[2]{Eqs.~(\ref{#1})--(\ref{#2})}

\newcommand\pa{\partial}
\newcommand\pdif[2]{\frac{\pa #1}{\pa #2}}
                        
\newcommand\ee{\end{equation}}
\newcommand\be{\begin{equation}}
\newcommand\eea{\end{eqnarray}}
\newcommand\bea{\begin{eqnarray}}
\renewcommand{\topfraction}{0.99}

\twocolumn[\hsize\textwidth\columnwidth\hsize\csname 
@twocolumnfalse\endcsname

\title{Super-horizon perturbations and preheating}
\author{Andrew R.~Liddle,$^1$ David H.~Lyth,$^2$ Karim A.~Malik$^3$ and 
David Wands$^3$}
\address{$^1$Astrophysics Group, The Blackett Laboratory, 
Imperial College, Prince Consort Road, London SW7 2BZ, Great Britain} 
\address{$^2$School of Physics and Materials, University of Lancaster,
Lancaster LA1 4YB, Great Britain}
\address{$^3$Relativity and Cosmology Group, 
School of Computer Science and Mathematics,
University of Portsmouth,\\Portsmouth PO1 2EG, Great Britain}
\maketitle
\begin{abstract}
It has recently been claimed by Bassett et al.~that preheating after
inflation may affect the amplitude of curvature perturbations on large
scales, undermining the usual inflationary prediction. We analyze the
simplest model, and confirm the results of Jedamzik and Sigl and of
Ivanov that in linear perturbation theory the effect is
negligible. However the dominant effect is second-order in the field
perturbation and we show that this too is negligible,
and hence conclude that preheating has no significant influence on
large-scale perturbations in this model. We briefly discuss the
likelihood of an effect in other models.
\end{abstract}

\pacs{PACS numbers: 98.80.Cq \hfill hep-ph/9912473}

\vskip2pc]

\section{Introduction}

The standard inflationary paradigm is an extremely successful model in
explaining observed structures in the Universe (see
Refs.~\cite{LL,David+Tony} for reviews). The inhomogeneities originate
from the quantum fluctuations of the inflaton field, which on being
stretched to large scales become classical perturbations. The field
inhomogeneities generate a perturbation in the curvature of comoving
hypersurfaces, and later on these inhomogeneities are inherited by
matter and radiation when the inflaton field decays. In the simplest
scenario, the curvature perturbation on scales much larger than the
Hubble length is constant, and in particular is unchanged during the
inflaton decay.  This enables a prediction of the present-day
perturbations which does not depend on the specific cosmological
evolution between the late stages of inflation and the recent past
(say, before nucleosynthesis).

It has recently been claimed~\cite{Betal} that this simple picture may
be violated if inflation ends with a period of preheating, a violent
decay of the inflaton particles into another field (or even into
quanta of the inflaton field itself).  Such a phenomenon would
completely undermine the usual inflationary picture, and indeed the
original claim was that large-scale perturbations would be amplified
into the non-linear regime, placing them in conflict with observations
such as measurements of microwave background anisotropies. Given the
observational successes of the standard picture, these claims demand
attention.

In a companion paper \cite{separate}, we discuss the general criteria
under which large-scale curvature perturbations can vary. As has been
known for some time, this is possible provided there exist large-scale
non-adiabatic pressure perturbations, as can happen for example in
multi-field inflation models~\cite{modes,GBW,ss,David+Tony}. Under those
circumstances a significant effect is possible during preheating,
though there is nothing special about the preheating era in this respect and
this effect always needs to be considered in any multi-component
inflation model.

In this paper we perform an analysis of the simplest preheating model,
as discussed in Ref.~\cite{Betal}. We identify two possible sources of
variation of the curvature perturbation. One comes from large-scale
isocurvature perturbations in the preheating field into which the
inflaton decays; we concur with the recent analyses of Jedamzik and
Sigl \cite{jedam} and Ivanov \cite{ivan} that this effect is
negligible due to the rapid decay of the background value of the
preheating field during inflation. However, we also show that in fact
a different mechanism gives the dominant contribution, which is
second-order in the field perturbations coming from short-wavelength
fluctuations in the fields.  Nevertheless, we show too that this
effect is completely negligible, and hence that preheating in this
model has no significant effect on large-scale curvature
perturbations.

\section{Perturbation evolution}

An adiabatic perturbation is one for which all perturbations $\delta
x$ share a common value for $\delta x/\dot{x}$, where $\dot{x}$ is the
time dependence of the background value of $x$.  If the Universe is
dominated by a single fluid with a definite equation of state, or by a
single scalar field whose perturbations start in the vacuum state,
then only adiabatic perturbations can be supported. If there is more
than one fluid, then the adiabatic condition is a special case, but
for instance is preserved if a single inflaton field subsequently
decays into several components.  However, perturbations in a second
field, for instance the one into which the inflaton decays during
preheating, typically violate the adiabatic condition.

We describe the perturbations via the curvature perturbation on
uniform-density hypersurfaces, denoted $\zeta$.\footnote{This is the
notation of Bardeen, Steinhardt and Turner~\cite{BST}. General issues
of perturbation description and evolution are discussed in a companion
paper \cite{separate}.  The curvature perturbation of comoving spatial
hypersurfaces, usually denoted by ${\cal R}$ \cite{LL,David+Tony}, is
practically the same as $\zeta$ well outside the horizon, since the
two coincide in the large-scale limit.}  In linear theory the
evolution of $\zeta$ is well known, and arises from the non-adiabatic
part of the pressure perturbations. In any gauge, the pressure
perturbation can be split into adiabatic and entropic (non-adiabatic)
parts, by writing
\begin{equation}
\delta p = c_{{\rm s}}^2 \delta\rho + \delta p_{\rm nad} \,,
\end{equation}
where $c_{\rm s}^2\equiv \dot p/\dot \rho$ and the non-adiabatic part is
\begin{equation}
\label{defGamma}
\delta p_{\rm nad}\equiv \dot p \Gamma \equiv \dot{p} \left( {\delta p
 \over \dot{p}} - {\delta\rho \over \dot{\rho}} \right) \,.
\end{equation}
The entropy perturbation $\Gamma$, defined in this way, is
gauge-invariant, and represents the displacement between hypersurfaces
of uniform pressure and uniform density.

On large scales anisotropic stress can be ignored when the matter
content is entirely in the form of scalar fields, and in its absence
the non-adiabatic pressure perturbation determines the variation of
$\zeta$, according to the equation \cite{GBW,David+Tony}
\begin{equation}
\label{dzetadN}
{d\zeta \over dN} = - 3H c_{{\rm s}}^2 \Gamma \,,
\end{equation}
where $N \equiv \ln a$ measures the integrated expansion and $H$ is
the Hubble parameter. The uniform-density hypersurfaces become
ill-defined if the density is not a strictly decreasing function along
worldlines between hypersurfaces of uniform density, and one might
worry that this undermines the above analysis. However we can equally
well derive this evolution equation in terms of the density
perturbation on spatially-flat hypersurfaces,
$\delta\rho_{\psi}\equiv-(d\rho/dN)\zeta$, which remains
well-defined. Spatially-flat hypersurfaces are automatically
separated by a uniform integrated expansion on large scales, so the
perturbed continuity equation in this gauge takes the particularly
simple form
\begin{equation}
\label{ddeltarhor}
{d\delta\rho_{\psi} \over dN} = -3(\delta\rho_{\psi}+\delta 
p_{{\psi}}) \,.
\end{equation}
{}From this one finds that $\delta\rho_\psi\propto d\rho/dN$ for adiabatic 
perturbations and hence again we recover constant value for $\zeta$. 
However it is clearly possible for entropy perturbations to cause a
change in $\zeta$ on arbitrarily large scales when the non-adiabatic
pressure perturbation is non-negligible.

\section{Preheating}

During inflation, the reheat field into which the inflaton field
decays possesses quantum fluctuations on small scales just like the
inflaton field itself. As these perturbations are uncorrelated with
those in the inflaton field, the adiabatic condition will not be
satisfied, and hence there is a possibility that $\zeta$ might vary on
large scales.  Only direct calculation can demonstrate whether the
effect might be significant, and we now compute this effect
in the simplest preheating model, as analyzed in
Ref.~\cite{Betal}. This is a chaotic inflation model with scalar field
potential
\begin{equation}
V(\phi,\chi) = \frac{1}{2} \, m^2\phi^2 +  \frac{1}{2} \, g^2\phi^2\chi^2 \,,
\end{equation}
where $\phi$ is the inflaton and $\chi$ the reheat field.
Slow-roll inflation proceeds with $\phi \gtrsim m_{{\rm Pl}}$ and
$g\chi\ll m$. The effective mass of the $\chi$ field is $g \phi$ and thus
will be much larger than the Hubble rate,
$H \simeq \sqrt{4\pi/3} \, m \phi/m_{{\rm Pl}}$, 
for $g \gg m/m_{{\rm Pl}} \sim 10^{-6}$. Throughout, we use the symbol 
`$\simeq$' to indicate equality within the slow-roll approximation.

This model gives efficient preheating, since the effective mass of 
$\chi$ oscillates about zero with large amplitude.
In most other models of inflation, preheating is less 
efficient or absent, because the mass oscillates about a nonzero
value and/or has a small amplitude.

Any variation of $\zeta$ during preheating will be driven by the
(non-adiabatic part of) the $\chi$ field perturbation. Our calculation
takes place in three steps. The first is to compute the perturbations
in the $\chi$ field at the end of inflation. The second is to compute
how these perturbations are amplified during the preheating epoch by
the strong resonance. Finally, the main part of the calculation is to
compute the change in $\zeta$ driven by these $\chi$ perturbations.

\subsection{The initial quantum fluctuation of the $\chi$-field}

Perturbations in the $\chi$ field obey the wave equation
\begin{equation}
\label{wavechi} 
\ddot{\delta\chi} + 3H\dot{\delta\chi} + 
\left(\frac{k^2}{a^2}+g^2\phi^2\right)\delta\chi =0 \,.
\end{equation}
The slow-roll conditions ensure that the $\chi$ field remains in the 
adiabatic vacuum state for a massive field
\begin{equation}
\delta\chi_k \simeq {e^{-i\omega t} \over \sqrt{2\omega}} \,,
\end{equation}
where $\omega^2=k^2/a^2+g^2\phi^2$. This is a solution provided 
\begin{equation}
\nu \equiv \frac{m_{\chi}}{H} \simeq \sqrt{\frac{3}{4\pi}}
        \frac{g\,m_{{\rm Pl}}}{m} \gg 1 \,,
\end{equation}
where $m_\chi \equiv g\phi$ is the effective mass of the $\chi$ field.

The power spectrum of a quantity $x$, decomposed into Fourier components 
$x_{{\bf k}}$, is defined as
\begin{equation}
{\cal P}_x \equiv \frac{k^3}{2 \pi^2} \left\langle|x_{{\bf k}}|^2
        \right\rangle \,,
\end{equation}
where $k = |{\bf k}|$ and the average is over ensembles. 
Hence the power spectrum for long-wavelength fluctuations ($k\ll
m_\chi$) in the $\chi$ field simply reduces to the result for a 
massive field in flat space 
\begin{equation}
\label{powspec1}
{\cal P}_{\delta\chi} \simeq \frac{1}{4\pi^2m_\chi}
        \left(\frac{k}{a}\right)^3 \,,
\end{equation}
where $m_\chi$ is the mass of the field at the required time.
Physically, this says that at all times the expansion of the Universe
has a negligible effect on the modes as compared to the mass.  In
particular, at the end of inflation we can write
\begin{equation}
\label{powspec2}
\left. {\cal P}_{\delta\chi} \right|_{{\rm end}} \simeq \frac{1}{\nu}
        \left(\frac{H_{{\rm end}}}{2\pi}\right)^2
        \left(\frac{k}{k_{{\rm end}}}\right)^3 \,.
\end{equation}
The power spectrum has a spectral index
$n_{_{\delta\chi}} = 3$. This is the extreme limit of the mechanism
used to give a blue tilt in isocurvature inflation scenarios \cite{LM}.

\subsection{Parametric resonance}

After inflation, the inflaton field $\phi$ oscillates. 
Strong parametric resonance may now occur, amplifying the initial 
quantum fluctuation in $\chi$ to become a perturbation of the classical 
field $\chi$. The condition for this is
\begin{equation}
q \equiv \frac{g^2 \Phi^2}{4m^2} \gg 1 \,,
\end{equation}
where $\Phi$ is the initial amplitude of the $\phi$-field oscillations.  

We model the effect of preheating on the amplitude
of the $\chi$ field following Ref.~\cite{KLS97} as
\begin{equation}
\label{chievol}
{\cal P}_{\delta\chi} =
\left. {\cal P}_{\delta\chi} \right|_{{\rm end}} \; e^{2\mu_{k} m\Delta 
t} \,,
\end{equation}
and the Floquet index $\mu_k$ is taken as 
\begin{equation}
\label{defmuk}
\mu_k\simeq\frac{1}{2\pi} 
        \ln \left( 1+ 2e^{-\pi\kappa^2} \right) 
\,,
\end{equation}
with
\begin{equation}
\label{defkappa}
\kappa^2
\equiv \left(\frac{k}{k_{{\rm max}}}\right)^2
\equiv \frac{1}{18\sqrt{q}}\left(\frac{k}{k_{{\rm end}}}\right)^2 
 \,.
\end{equation}
For strong coupling ($q \gg 1$), we have $\kappa^2\ll1$ for
all modes outside the Hubble scale after inflation
ends $(k \le k_{{\rm end}})$.  Therefore $\mu_k\approx
\ln3/2\pi\approx 0.17$ is only very weakly dependent on the wavenumber
$k$. Combining Eqs.~(\ref{powspec2}) and (\ref{chievol}) gives
\begin{equation}
\label{powspec3}
{\cal P}_{\delta\chi} \simeq {1\over \nu} \left(\frac{H_{{\rm
        end}}}{2\pi}\right)^2\left(\frac{k}{k_{{\rm end}}}\right)^3
        \; e^{2\mu_km\Delta t} \,.
\end{equation}

\subsection{Change in the curvature perturbation on large scales}

In order to quantify the effect parametric growth of the $\chi$ field
fluctuations during preheating might have upon the standard
predictions for the spectrum of density perturbations after inflation,
we need to estimate the change in the curvature perturbation $\zeta$
on super-horizon scales due to entropy perturbations on large-scales.

The density and pressure perturbations due to first-order
perturbations in the inflaton field on large scales (i.e.~neglecting
spatial gradient terms) are of order $g^2\phi^2\chi\delta\chi$. Not
only are the field perturbations $\delta\chi$ strongly suppressed on
large scales at the end of inflation [as shown in our
Eq.~(\ref{powspec2})] but so is the background field $\chi$.
We can place an upper bound on the size of the background field by
noting that in order to have slow-roll chaotic inflation (dominated by
the $m^2\phi^2/2$ potential) when any given mode $k$ which we are
interested in crossed outside the horizon, we require $\chi\ll
m/g$. The large effective mass causes this background field to decay,
just like the super-horizon perturbations, and at the end of inflation
we require $\chi\ll m/g(k/k_{\rm end})^{3/2}$ when considering
preheating in single-field chaotic inflation. Combining this with
Eq.~(\ref{powspec2}) we find that the spectrum of density or pressure
perturbations due linear perturbations in the $\chi$ field has an
enormous suppression for $k\ll k_{\rm end}$:
\begin{equation}
\left. {\cal P}_{\chi\delta\chi} \right|_{\rm end} \ll 
\sqrt{{4\pi\over3}} \left( {m\over gm_{\rm Pl}} \right)^3 
\left( {m_{\rm Pl}H_{\rm end} \over 2\pi} \right)^2 
\left( {k\over k_{\rm end}} \right)^6 
\end{equation}
Effectively the density and pressure perturbations have no term linear
in $\delta \chi$, because that term is multiplied by the background
field value which is vanishingly small. 

By contrast the second-order pressure perturbation is of order
$g^2\phi^2\delta\chi^2$ where the power spectrum of $\delta\chi^2$ is
given by~\cite{axion}
\begin{equation}
\label{sqrspec}
{\cal P}_{\delta\chi^2} \simeq {k^3\over2\pi} \int_0^{k_{\rm cut}}
\frac{{\cal P}_{\delta\chi}(\left|{\bf k}'\right|)
{\cal P}_{\delta\chi}(\left|{\bf k}-{\bf k}'\right|)}
{\left|{\bf k}\right|^{\prime3}\left|{\bf k}-{\bf k}' \right|^3} 
d^3{\bf k}' \,.
\end{equation}
We impose the upper limit $k_{\rm cut}\sim k_{\rm max}$ to 
eliminate the ultraviolet divergence associated with vacuum state.
Substituting in for ${\cal P}_{\delta\chi}$ from
Eq.~(\ref{powspec2}), we can write
\begin{equation}
\left. {\cal P}_{\delta\chi^2} \right|_{\rm end}
 = {8\pi\over9} \left( {m\over gm_{\rm Pl}} \right)^2 
\left( {H_{\rm end} \over
2\pi} \right)^4 \left( {k_{\rm cut} \over k_{\rm end}} \right)^3 
\left( {k \over k_{\rm end}} \right)^3 \,,
\end{equation}
Noting that $H_{\rm end}\sim m$ and $k_{\rm cut}\sim k_{\rm max}\sim
q^{1/4}k_{\rm end}$, it is evident that the second-order effect will
dominate over the linear term for $k<g^{1/2}q^{1/3}k_{\rm
end}$.

The leading-order contributions to the pressure and density
perturbations on large scales are thus
\begin{eqnarray}
\label{deltarho}
\delta\rho &=& m^2\phi\,\delta\phi + \dot\phi\,\dot{\delta\phi}
 +{1\over2} g^2\phi^2\delta\chi^2 + {1\over2}\dot{\delta\chi}^2 \,,\\
\label{deltap}
\delta p &=& - m^2\phi\,\delta\phi + \dot\phi\,\dot{\delta\phi}
 - {1\over2} g^2\phi^2\delta\chi^2 + {1\over2}\dot{\delta\chi}^2 \,.
\end{eqnarray}
We stress that we will still only consider first-order perturbations
in the metric and total density and pressure, but these include
terms to second-order in $\delta\chi$.
{}From Eqs.~(\ref{defGamma}), (\ref{deltarho}) and~(\ref{deltap}) we
obtain
\begin{equation}
\delta p_{\rm nad} =
\frac {-m^2\phi\,\dot{\delta\chi}^2 + \ddot\phi
g^2\phi^2\delta\chi^2}{3H\dot\phi}
\,,
\end{equation}
where the long-wavelength solutions for vacuum fluctuations in the
$\phi$ field obey the adiabatic condition
$\delta\phi/\dot\phi=\dot{\delta\phi}/\ddot\phi$.  Inserted into
\eq{dzetadN}, this gives the rate of change of $\zeta$.

Note that the non-adiabatic pressure will diverge periodically when
$\dot\phi=0$ as the comoving or uniform density hypersurfaces become
ill-defined.  Such a phenomenon was noted in the single-field context
by Finelli and Brandenberger \cite{FB}, who evaded it by instead using
Mukhanov's variable $u=a\delta\phi_{\psi}$ which renders well-behaved
equations. Linear perturbation
theory remains valid as there are choices of hypersurface, such as the
spatially-flat hypersurfaces, on which the total pressure perturbation
remains finite and small.  In particular, we can calculate the change
in the density perturbation due to the non-adiabatic part of the
pressure perturbation on spatially-flat hypersurfaces from
Eq.~(\ref{ddeltarhor}), which yields
\begin{equation}
\Delta\rho_{\rm nad} = -3\int \delta p_{\rm nad} H dt \, .
\end{equation}
Even though $\delta p_{\rm nad}$ contains poles whenever $\dot\phi=0$,
the integrated effect remains finite whenever the upper and lower limits
of the integral are at $\dot\phi\neq0$.
{}From this density perturbation calculated in the spatially-flat gauge
one can reconstruct the change in the curvature perturbation on uniform
density hypersurfaces
\begin{equation}
\Delta\zeta = - H {\Delta\rho_{\rm nad} \over \dot\rho} \,.
\end{equation}

Substituting in our expression for $\delta p_{\rm nad}$ we obtain
\begin{equation}
\label{Deltazeta}
\Delta\zeta = {1\over \dot\phi^2} \int \left( 1+ {2m^2\phi\over3H\dot\phi}
\right) g^2\phi^2 \left|\delta\chi^2\right| H\, dt \,,
\end{equation}
where we have averaged over short timescale oscillations of the
$\chi$-field fluctuations to write $\left|\dot{\delta\chi}^2\right| =
g^2\phi^2 \left|\delta\chi^2\right|$.
To evaluate this we take the usual adiabatic evolution for the
background $\phi$ field after the end of inflation
\begin{equation}
\phi = \Phi \, {\sin(m\Delta t) \over m\Delta t} \,,
\end{equation}
and time-averaged Hubble expansion
\begin{equation}
H = {2m \over 3(m\Delta t + \Theta)} \,,
\end{equation}
where $\Theta$ is an integration constant of order unity.
The amplitude of the $\chi$-field fluctuations also decays proportional
to $1/\Delta t$ over a half-oscillation from $m\Delta t=n\pi$ to
$m\Delta t=(n+1)\pi$, with the stochastic growth in particle
number occurring only when $\phi=0$.
Thus evaluating $\Delta\zeta$ over a half-oscillation $\Delta t=\pi/m$
we can write
\begin{equation}
\Delta\zeta = {2g^2 |\delta\chi^2| x_n^4 \over 3m^2} 
 \int_{x_n}^{x_{n+1}} \left( {1\over x+\Theta} + {s\over s'} \right)
{s^2\over x^2} dx \,,
\end{equation}
where $x=m\Delta t$, $s(x)=\sin x/x$, $x_n=n\pi$ and a dash indicates
differentiation with respect to $x$. The integral is dominated by the
second term in the bracket which has a pole of order 3 when $s'=0$.
Although $s/s'$ diverges, it yields a finite contribution to the
integral which can be evaluated numerically. For $x_n\gg1$ the integral
is very well approximated by $24/x_n^4$, independent of the integration
constant $\Theta$.

This expression gives us the rate of change of the curvature
perturbation $\zeta$ due to the pressure of the field fluctuations
$\delta\chi^2$ over each half-oscillation of the inflaton field $\phi$. 
Approximating the sum over several oscillations as a smooth integral and
using Eq.~(\ref{chievol}) for the growth of the $\chi$-field fluctuations
during preheating (neglecting the weak $k$-dependence of the Floquet
index, $\mu_k$, on super-horizon scales) we obtain
\begin{equation}
\label{zetanad}
\zeta_{\rm nad} = {16 g^2 \over 2\pi\mu} {\left|\delta\chi^2\right|
_{\rm end} \over m^2} e^{2\mu m\Delta t} \,.
\end{equation}

The statistics of these second-order fluctuations are non-Gaussian,
being a $\chi^2$-distribution. Both the mean and the variance of
$\zeta_{{\rm nad}}$ are non-vanishing. The mean value will not
contribute to density fluctuations, but rather indicates that the
background we are expanding around is unstable as energy is
systematically drained from the inflaton field.  We are interested in
the variance of the curvature perturbation, and in particular the
change of the curvature perturbation power spectrum on super-horizon
scales which is negligible if the power spectrum of $\zeta_{{\rm
nad}}$ on those scales is much less than that of $\zeta$ generated
during inflation, the latter being required to be of order $10^{-10}$
to explain the COBE observations.

To evaluate the power spectrum for $\zeta_{\rm nad}$ we must evaluate
the power spectrum of $\delta\chi^2$ which is given 
by substituting ${\cal P}_{\delta\chi}$, from
Eq.~(\ref{powspec3}), 
into Eq.~(\ref{sqrspec}). This gives
\begin{equation}
{\cal P}_{\delta\chi^2} = {2\over3\nu^2} \left( {H_{\rm end} \over
2\pi} \right)^4 \left( {k_{\rm max} \over k_{\rm end}} \right)^3 
\left( {k \over k_{\rm end}} \right)^3 I(\kappa,m\Delta t)\,,
\end{equation}
where 
\begin{equation}
\label{integral}
I(\kappa,m\Delta t)\equiv \frac{3}{2} \int_0^{\kappa_{\rm cut}} \!\! d\kappa'
\int_0^\pi \! d\theta \, e^{2(\mu_{\kappa'}+\mu_{\kappa-\kappa'})m\Delta t} 
\kappa'^2 \sin\theta \,,
\end{equation}
$\kappa= k/k_{\rm max}$ as defined in Eq.~(\ref{defkappa}), 
and $\theta$ is the angle between ${\bf k}$ and ${\bf k}'$. 
Note that at the end of inflation we have $I(\kappa,0)=\kappa_{\rm
cut}^3\sim 1$, and ${\cal P}_{\delta\chi^2}\propto k^3$. 
This yields
\begin{equation}
\label{Pzetanad}
{\cal P}_{\zeta_{\rm nad}} 
\simeq {2^{9/2}3 \over \pi^5\mu^2} \left({\Phi \over m_{\rm Pl}}\right)^2 
\left({H_{\rm end} \over m}\right)^4 g^4 q^{-1/4} \left({k\over k_{\rm
end}}\right)^3 I \,.
\end{equation}

One might have thought that the dominant contribution to $\zeta_{{\rm
nad}}$ on large scales would come from $\delta\chi$ fluctuations on
those scales, and that is indeed the presumption of the calculation of
Bassett et al.~\cite{Betal}.  However, in fact the integral is
initially dominated by $k' \sim k_{{\rm cut}}$, namely the shortest
scales. The reason for this is the steep slope of ${\cal
P}_{\delta\chi}$; were it much shallower (spectral index less than
3/2), then the dominant contribution would come from large scales.

To study the scale dependence of $I(\kappa,m\Delta t)$ and hence
${\cal P}_{\zeta_{\rm nad}}$ at later times, we can
expand $\mu_{\kappa-\kappa'}$ for $\kappa \kappa'\ll 1$ as 
\begin{equation}
\mu_{\kappa-\kappa'} = \mu_{\kappa'} 
+\frac{2\kappa'\cos\theta}{2+e^{\pi \kappa'^2}}\, 
\kappa + {\cal O}(\kappa^2)\,.
\end{equation}
We can then write the integral in Eq.~(\ref{integral}) as
\begin{equation}
I(\kappa,m\Delta t)
 = I_0(m\Delta t)+ {\cal O}(\kappa^2)\,,
\end{equation}
where first-order terms, ${\cal O}(\kappa)$, vanish by symmetry and
\begin{equation}
I_0(m\Delta t)=\frac{3}{2}
\int_0^{\kappa_{\rm cut}} e^{4 \mu_{\kappa'} m \Delta t}
\kappa'^2 d\kappa' \,.
\end{equation}
Thus the scale dependence of ${\cal P}_{\zeta_{\rm nad}}$ remains $k^3$
on large-scales for which $\kappa\ll 1$.

At late times these integrals become dominated by the 
modes with $\kappa'^2\ll (m\Delta t)^{-1}$ which are preferentially
amplified during preheating. These are longer wavelength 
than $k_{{\rm cut}}$, but still very short compared to the scales 
which give rise to large scale structure in the present Universe.
{}From Eq.~(\ref{defmuk}) we have
$\mu_{\kappa'} \approx \mu_0 - \kappa'^2/3$, for $\kappa'^2\ll1$, where
$\mu_0=(\ln 3)/2\pi$, which gives the asymptotic behaviour at late
times
\begin{equation}
I_0 \simeq 0.86 (m\Delta t)^{-3/2}e^{4\mu_0 m\Delta t} \,.
\end{equation}
Thus although the rate of growth of ${\cal P}_{\zeta_{\rm nad}}$
becomes determined by the exponential growth of the long-wavelength
modes, the scale dependence on super-horizon scales remains
proportional to $k^3$ for $\kappa \lesssim (m\Delta t)^{-1/2}$. This
ensures that there can be no significant change in the curvature
perturbation, $\zeta$, on very large scales before back-reaction on
smaller scales becomes important and this phase of preheating
ends when $m\Delta t\sim100$~\cite{KLS97}.

\begin{figure}[t]
\centering 
\leavevmode\epsfysize=6cm \epsfbox{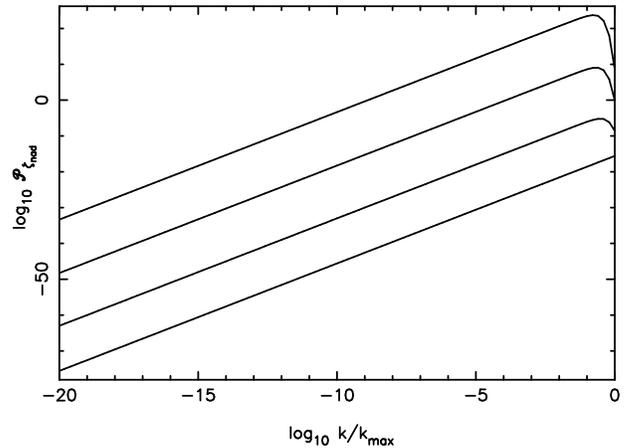}\\
\caption[specint]{\label{specint} The power spectrum of the
non-adiabatic curvature perturbation ${\cal P}_{\zeta_{{\rm nad}}}$,
shown at four different times: from bottom to top $m\Delta t = 0$,
$50$, $100$ and $150$. The parameters used were $g=10^{-3}$,
$m=10^{-6} m_{{\rm Pl}}$ and $k_{\rm cut}=k_{\rm max}$.}
\end{figure} 

Numerical evaluation of Eq.~(\ref{Pzetanad}) confirms our analytical
results, as shown in Fig.~1. For $k \ll k_{{\rm max}}$, the spectral
index remains $k^3$ during preheating. Observable scales have
$\log_{10} k/k_{{\rm max}} \simeq -20$.

Our result shows that because of the $k^3$ spectrum of $\delta \chi$,
which leads to a similarly steep spectrum for $\zeta_{{\rm nad}}$,
there is a negligible effect on the large-scale perturbations before
the resonance ceases.  The suppression of the large-scale
perturbations in $\delta\chi$, discussed in Refs.~\cite{jedam,ivan},
means that large-scale perturbations in $\delta\chi$ are completely
unimportant. However, it turns out that they don't give the largest
effect, which comes from the short-scale modes which dominate the
integral for $\zeta_{{\rm nad}}$. Nevertheless, even they give a
negligible effect, again with a $k^3$ spectrum. Indeed, that result
with hindsight can be seen as inevitable; it has long been known
\cite{causal} that local processes conserving energy and momentum
cannot generate a tail shallower than $k^3$ (with our spectral index
convention) to large scales, which is the Fourier equivalent of
realizing that in real space there is an upper limit to how far energy
can be transported. Any mechanism that relies on short-scale
phenomena, rather than acting on pre-existing large-scale
perturbations, is doomed to be negligible on large scales.

\section{Conclusions and discussion}

As discussed in detail in a companion paper \cite{separate},
large-scale curvature perturbations can vary provided there is a
significant non-adiabatic pressure perturbation. This is always
possible in principle if there is more than one field or fluid, and
since preheating usually involves at least one additional field into
which the inflaton resonantly decays, such variation is in principle
possible.

In this paper we have focussed on the simplest preheating model, as
discussed in Ref.~\cite{Betal}. We have identified the non-adiabatic
pressure, and shown that the dominant effect comes from second-order
perturbations in the preheating field.  Further, the effect is
dominated by perturbations on short scales, rather than from the
resonant amplification of non-adiabatic perturbations on the large
astrophysical scales. Nevertheless, we have shown that the
contribution has a $k^3$ spectrum to large scales, rendering it
totally negligible on scales relevant for structure formation in our
present Universe by the time backreaction ends the resonance. Amongst
models of inflation involving a single-component inflaton field, this
model gives the most preheating, and so this negative conclusion will
apply to all such models.

Recently Bassett et al.~\cite{Betal2} have suggested large effects
might be possible in more complicated models. They consider two types
of model. In one kind, inflation takes place along a steep-sided
valley, which lies mainly along the direction of a field $\phi$ but
with a small component along another direction $\chi$. In this case,
one can simply define the inflaton to be the field evolving along the
valley floor, and the second heavy field lies orthogonal to it.
Taking that view, there is no reason to expect the preheating of the
heavy field to give rise to a bigger effect than in the simpler model
considered in this paper.

In the second kind of model, the reheat field is light during
inflation, and this corresponds to a two-component inflaton field. As
has long been known, there can indeed be a large variation of $\zeta$
in this case, which can continue until a thermalized
radiation-dominated universe has been established. Indeed, in models
where one of the fields survives to the present Universe (for example
becoming the cold dark matter), variation in $\zeta$ can continue
right to the present. This variation is due to the presence on large
scales of classical perturbations in both fields (properly thought of
as a multi-component inflaton field) generated during inflation, and
the effect of these must always be considered in a multi-component
inflation model, with or without preheating.

\section*{Acknowledgments}

We thank Bruce Bassett, Lev Kofman, Andrei Linde, Roy Maartens and Anupam 
Mazumdar for useful discussions. DW is supported by the Royal Society.

 
\end{document}